\documentclass[twocolumn,preprintnumbers,amsmath,amssymb]{revtex4-1}
\usepackage{graphicx}
\usepackage{ulem}
\usepackage[svgnames]{xcolor}
\usepackage{bm}
\usepackage{multirow}

\newif\ifshowcomments\showcommentstrue

\begin{document}

\title{
Chiral domain walls of Mn$_3$Sn and their memory }

\author{  Xiaokang Li$^{1,2}$, Cl\'ement Collignon$^{2,3}$, Liangcai Xu$^{1}$, Huakun Zuo$^{1}$,  Antonella Cavanna$^{4}$,  Ulf Gennser$^{4}$, Dominique Mailly$^{4}$, Beno\^{\i}t Fauqu\'e$^{2,3}$, Leon Balents$^{5}$, Zengwei Zhu$^{1,*}$ and Kamran Behnia$^{2,6,*}$}

\affiliation{(1) Wuhan National High Magnetic Field Center and School of Physics, Huazhong University of Science and Technology,  Wuhan  430074, China\\
 (2) Laboratoire de Physique et d'Etude de Mat\'{e}riaux (CNRS)\\ ESPCI Paris, PSL Research University, 75005 Paris, France\\
(3) JEIP  (USR 3573 CNRS), Coll\`ege de France,  75005 Paris, France\\
(4) Centre de Nanosciences et de Nanotechnologies, CNRS,  Universit\'e Paris-Saclay, 91120 Palaiseau, France\\
(5) Kavli Institute for Theoretical Physics, University of California Santa Barbara, California 93106, USA\\
(6) II. Physikalisches Institut, Universit\"{a}t zu K\"{o}ln, 50937 K\"{o}ln, Germany
}

\date{February 28, 2019}
\begin{abstract}
Magnetic domain walls are topological solitons whose internal structure is set by competing energies which sculpt them. In common ferromagnets, domain walls are known to be of either Bloch or N\'eel types. Little is established in the case of Mn$_3$Sn, a triangular antiferromagnet with a large room-temperature anomalous Hall effect, where  domain nucleation is triggered by a well-defined threshold magnetic field. Here, we show that the domain walls of this system generate an additional contribution to the Hall conductivity tensor and a transverse magnetization. The former is an electric field lying in the same plane with the magnetic field and electric current and therefore a planar Hall effect.  We demonstrate that in-plane rotation of spins inside the domain wall would explain both observations and the clockwise or anticlockwise chirality of the walls depends on the history of the field orientation and can be controlled.
\end{abstract}

\maketitle
\section{Introduction}
A domain wall is the topological defect of a discrete symmetry. In ferromagnetic materials, these are narrow boundaries separating magnetic domains with different polarities. Their width and structure are  set by the competition between the exchange energy and the magneto-crystalline anisotropy energy \cite{Getzlaff}. They are either of Bloch type, where the the magnetization rotates in a plane parallel to the wall plane, or of N\'eel type, whose magnetization vector rotates in a plane perpendicular to the wall. Thanks to high-resolution scanning probes of local magnetization, they can be visualized \cite{Tetienne}. Theoretical proposals for other more sophisticated spin textures have recently emerged\cite{Cheng}. In addition to their fundamental interest, the attention to domain walls is driven by the quest for new spintronic devices \cite{Stamps}. Much less is known about antiferromagnetic domain walls.

A large anomalous Hall effect (AHE) was recently discovered \cite{Nakatsuji,Nayak,Kiyohara} in the Mn$_3$X(X=Sn,Ge) family of non-collinear antiferromagnets \cite{Zimmer,Tomiyoshi1982,Tomiyoshi1982b,Yang2017}. The discovery followed theoretical predictions \cite{Chen2014,Kubler2014} and preceded the observation of  a variety of other anomalous transverse responses by thermal and optical probes  \cite{Ikhlas2017,Li2017,Higo2018,Xu2018,Balk2019,Wuttke2019,Sugii2019}. These materials constitute new platforms for antiferromagnetic spintronics \cite{Zhang2017,Smejkal2018}. The structure of domain walls have been a subject of theoretical \cite{Liu2017} and experimental studies \cite{Li2018}.  Evidence and arguments for a non-trivial spin texture in domain walls are available, but no direct image of their magnetic structure, yet.

Here, we report on three distinct experimental observations leading us to identify the in-plane structure of the domain walls in  Mn$_3$Sn. The first observation is that in the narrow magnetic field window of multiple domains, there is a planar Hall effect (PHE) which consists in an electric field oriented parallel (and not perpendicular) to the applied magnetic field. The thermoelectric counterpart of this effect, namely a planar Nernst effect (PNE) was also detected. The second observation is the existence of a transverse magnetic response in the same narrow field window. Employing micron-size Hall sensors in close proximity with the sample\cite{Behnia2000,Collignon2017}, we monitored the local magnetic field at the surface and found in the same field window  a finite off-diagonal magnetization: a finite magnetization oriented perpendicular to the orientation of the applied magnetic field. We will argue below that a satisfactory explanation of both these observations is provided by a specific spin texture inside the domain walls. The third result is that the sign of the emergent electric field (set by the clock-wise or anti-clockwise rotation of the spins inside walls) depends on the history of the magnetic field orientation. We will show that this is caused by residual minority domains promoting a specific chirality. This last observation constitutes a new case of memory formation in condensed matter recording a direction \cite {Keim2018}.

\begin{figure}
\includegraphics[width=9cm]{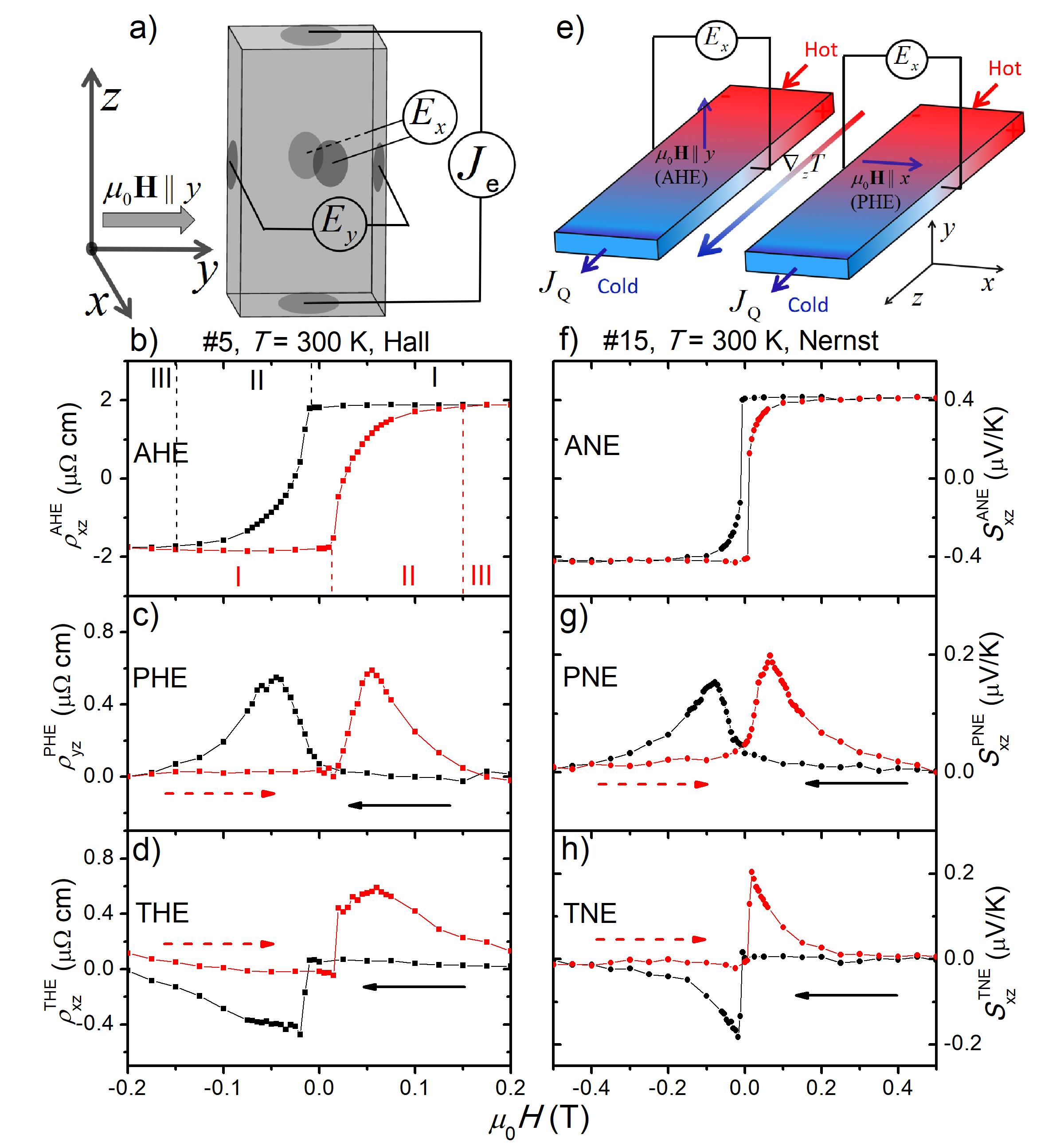}
\caption{\textbf{Room-temperature anomalous transverse response:} (a) Experimental configuration for measuring Hall effect in sample \#5 with square cross section.  Charge current is applied along the z-axis and the magnetic field along the y-axis. Two pairs of electrodes  measure $E_x$ and $E_y$.  (b) Anomalous Hall resistivity ($\rho^{\rm{AHE}}_{xz}$), extracted from $E_x$. (c) Planar Hall resistivity ($\rho^{\rm{PHE}}_{yz}$), extracted from $E_y$. (d)  Topological Hall resistivity ($\rho^{\rm{THE}}_{xz}$), extracted by subtracting magnetization and Hall hysteresis loops, see Supplementary Note 4. (e) Experimental configuration for measuring Nernst effect in sample \#15 with rectangular cross section. The temperature gradient is applied along the $z$-axis, the magnetic field is applied either along $x$-axis or $y$-axis. The electric field is always measured along the $x$-axis. (f) Anomalous Nernst  effect ($S^{\rm{ANE}}_{xz}$) with the magnetic field along the y-axis, extracted from $E_x$. (g) Planar Nernst effect ($S^{\rm{PNE}}_{xz}$) measured with the magnetic field along $x$-axis, extracted from $E_x$. (h) Topological Nernst effect ($S^{\rm{TNE}}_{xz}$) extracted by subtracting magnetization and Nernst hysteresis loops, see Supplementary Note 4. The larger width of the hysteresis loop in the Nernst measurements is due to the larger aspect ratio of the sample (See Fig. 2b ).}
\label{fig:Room-Tem-PHE-PNE}
\end{figure}
\section{Results}
\subsection{Planar Hall effect and Planar Nernst effect}
Fig.~\ref{fig:Room-Tem-PHE-PNE} shows an additional hitherto unreported component in the  Hall  and the  Nernst responses of Mn$_3$Sn, which we call planar Hall effect (PHE) and planar Nernst effect (PNE). The experimental configuration is sketched in Fig.~\ref{fig:Room-Tem-PHE-PNE}(a). Charge current was applied along the $z$-axis ($J//z$) and the magnetic field was oriented along the  y-axis ($H//y$). Electric field was measured simultaneously along both $x$- and $y$-axes. $E_x$, which represents an electric field vector perpendicular the magnetic field and the charge current, is the Hall response. As seen in Fig.~\ref{fig:Room-Tem-PHE-PNE}(b), it displays a hysteretic jump as reported previously \cite{Nakatsuji, Li2017,Li2018}. As the magnetic field is swept, three different regimes succeed each other \cite{Li2018}. In regime I, the system hosts one single domain. When the applied magnetic field (opposite to  the magnetization of the dominant domain) exceeds a threshold, new domains nucleate and regime II starts. At sufficiently large magnetic field, the system becomes single-domain again (regime III).  As seen in (Fig.~\ref{fig:Room-Tem-PHE-PNE}c),  in regime II,  $E_y$, the component of the electric field parallel to the magnetic field, becomes finite. The result was reproduced in several other samples and was also detected when the applied magnetic field  was along the $x$-axis, see Supplementary Figure 2. In other words,in the presence of multiple domains, when  $J//z$ and $H//y (//x)$, there is a non-vanishing $E_y$ ($E_x$). This is a planar Hall effect, with an electric field, which is parallel and not perpendicular to the magnetic field.  Note that this signal only emerges in the presence of domain walls. Its amplitude is comparable to the amplitude of the topological Hall effect (THE) Fig.~\ref{fig:Room-Tem-PHE-PNE}(d) extracted by subtracting Hall and magnetization hysteresis loops \cite{Li2018}, see Supplementary Note 4. Interestingly, the THE is present in the same field interval as the PHE, but shows different signs for the two sweeping orientations.

The experimental configuration for probing the Nernst response is shown in Fig.~\ref{fig:Room-Tem-PHE-PNE}e. The thermal gradient is applied along the $z$-axis. When the  magnetic field is oriented along the $y$-axis, there is a finite $E_x$. It represents the anomalous Nernst effect, which also displays a hysteretic jump (Fig.~\ref{fig:Room-Tem-PHE-PNE}f), as reported previously \cite{Ikhlas2017,Li2017}. In addition to this, however,  when the magnetic field is along the $x$-axis, there is  a finite $E_x$ in regime II (Fig.~\ref{fig:Room-Tem-PHE-PNE}g). This is the planar Nernst effect (PNE). Like its Hall counterpart, it becomes non-zero in a narrow field window when there are multiple domains and its amplitude is comparable to the amplitude of the topological Nernst effect (TNE)  (Fig.~\ref{fig:Room-Tem-PHE-PNE}h) extracted by subtracting Nernst and magnetization hysteresis loops, see Supplementary Note 4.


We carried out an extensive set of temperature-dependent measurements, see Supplementary Figure 4.  In the whole temperature window of the triangular order in Mn$_3$Sn ($50 K<T< 300 K$), the magnitude of PHE (PNE) remain a sizable fraction ($\approx 0.3-0.4$) of the total AHE (ANE) and there is no significant evolution with temperature. We will show  below how the PHE, the PNE and their odd parity in field, are set by the internal structure of domain walls \cite{Liu2017} in this system.

\begin{figure}
\includegraphics[width=9cm]{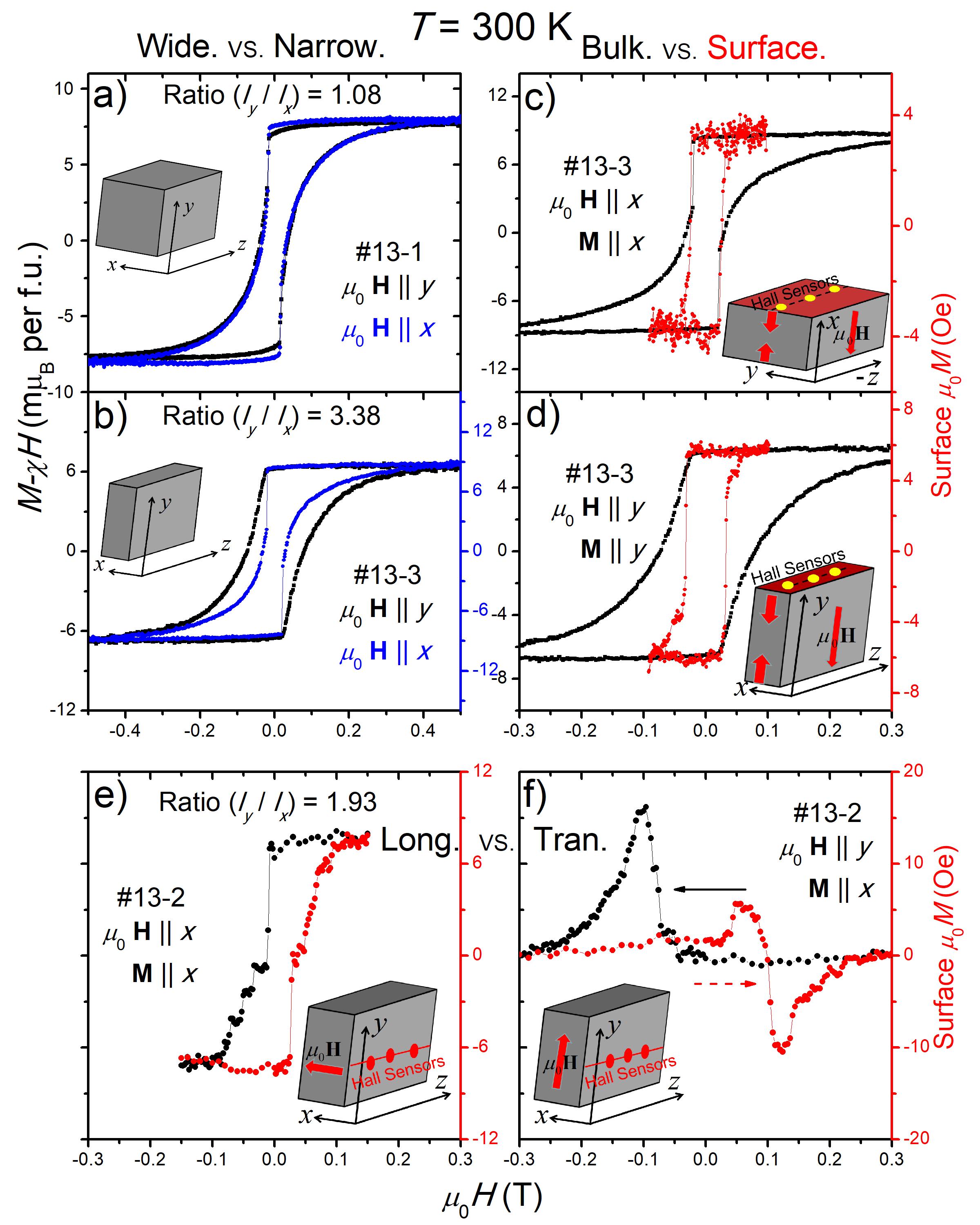}
\caption{\textbf{Bulk and surface magnetization:} (a)  Bulk magnetization (after subtracting the high-field slope) in a sample with an almost square cross-section  ($l_y/l_x=1.08$) for two field orientations. (b) Same in a sample  where ($l_y/l_x=3.38$). The hysteresis loop is wider when the field is oriented along the longer axis of the sample.  (c) Bulk and surface magnetization in sample \#13-3 for field along $x$. (d) Bulk and surface magnetization in sample \#13-3 for field along $y$. In contrast to  bulk magnetization, the hyteresis loop for surface magnetization is always narrow and does not depend on the aspect ratio. (e)  Longitudinal surface magnetization coefficient $\mu_{0}M_{xx}$; (f) transverse surface magnetization coefficient $\mu_{0}M_{yx}$ in sample \#13-2.
\label{fig:Bulk-surface-magn}}
\end{figure}

\subsection{Magnetization (bulk \textit{vs.} surface; longitudinal \textit{vs.} transverse)}
Fig.~\ref{fig:Bulk-surface-magn} presents the data magnetization obtained in two different ways. In addition to measuring bulk magnetization with a conventional vibrating sample magnetometer (VSM), we used two-dimensional electron gas (2DEG) Hall sensors, attached to one edge of the sample to monitor the local magnetic field at its surface (See method). By choosing the mutual orientation of the sensor and the applied magnetic field, we could extract both diagonal and  off-diagonal magnetization at the surface of the sample.

As seen in Fig.~\ref{fig:Bulk-surface-magn}a and Fig.~\ref{fig:Bulk-surface-magn}b, the hysteresis loop of bulk magnetization  depends on the aspect ratio $l_y/l_x$, where $l_y (l_x$) is the length of the sample along the $y$-axis($x$-axis). When $l_y/l_x \approx 1$  [inset.(a)],  bulk magnetization for the field along two orientations are almost coincident. But when $l_y/l_x \approx 3$, the hysteresis loop is wider when the field is oriented along the longer axis, in agreement with that was reported previously \cite{Li2017}. As seen in the Fig. \ref{fig:Bulk-surface-magn}a, domain nucleation occurs at the same magnetic field for the two orientations, but the loop closes later when the field is oriented along the longer axis. A straightforward interpretation of this observation is that the new domain(s) occupy the whole sample more efficiently when the magnetic field is oriented along a shorter axis.

Additional insight is brought by surface magnetization data obtained with Hall sensors. As shown in Fig.~\ref{fig:Bulk-surface-magn}c and Fig.~\ref{fig:Bulk-surface-magn}d, no matter the sample's aspect ratio, the hysteresis loop of surface magnetization is always narrow. The surface magnetization shows a sharp jump at the threshold field of bulk magnetization. We conclude that when the field is oriented along $y$- ($x$-) axis, the new domains nucleate at the $xz (yz)$- surface and immediately occupy the area ($5\times 5\mu m^2$) probed by a Hall sensor. The wide hysteresis loop of the bulk magnetization monitors the gradual enhancement produced by the smooth occupation of the center of the sample.


We used the Hall sensors to look for an off-diagonal magnetic response, namely a finite magnetic field perpendicular to the applied field. The mutual configuration of the sample, the magnetic field and the Hall sensors for quantifying longitudinal and transverse magnetization  are shown in [inset.(e)] and [inset.(f)]. The obtained data at room temperature is shown in Fig.~\ref{fig:Bulk-surface-magn}e and Fig.~\ref{fig:Bulk-surface-magn}f. The transverse response is restricted to regime II and has symmetric and asymmetric components.

\begin{figure*}
\centering
\includegraphics[width=18cm]{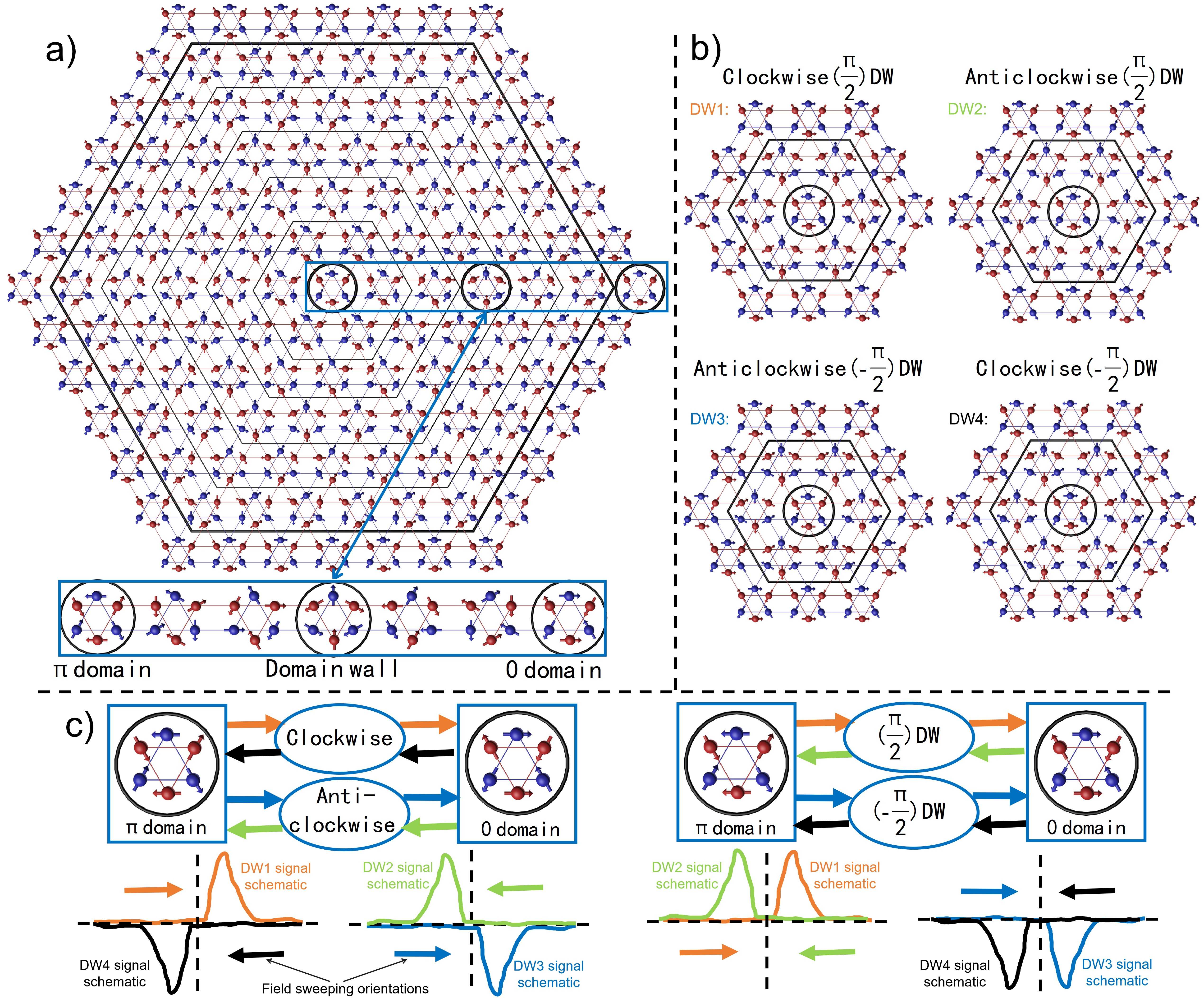}
\caption{\textbf{Spin texture in a domain wall and the signals they generate:}  (a)  Red and blue circles represent Mn atoms in adjacent planes. Arrows represent spins. A $\theta=\rm{\pi}$ domain (at the center), a $\theta=0$ domain (at the periphery) and the wall separating them (in between). For the sake of illustration, the thickness of the wall is assumed to be only five unit cells. Along a radial direction, adjacent equivalent spins  tilt by a constant angle.  A zoom on the wall along x-axis shows that the domain wall can has a $\theta=\pm \rm{\pi}/2$ spin configuration (b) Domain walls can have either $+\rm{\pi}/2$ or $-\rm{\pi}/2$ configurations. This could correspond to a clockwise or an anticlockwise rotation of spins as one moves from the center towards the periphery. (c) The orientation of the spins at the center of domain wall would explain the finite planar  Hall effect (PHE) and the transverse magnetization (TM). The sign of these signals (their odd parity or even parity with respect to magnetic field in a  hysteresis loop from a $\theta=\rm{\pi}$ domain to $\theta=0$ and back) depends on the chirality of the domain wall. The diagrams represent a field sweep (from one domain to another passing by a specific type of domain wall). In each case the four possible sequences are identified with a colored arrow. In the two plots sketched below each diagram, the same color is used to represent the expected curves for the PHE and TM response. The left-side diagram and plots refer to a case in which  rotation keeps the same sense (either clockwise or anticlockwise). As a consequence, the peaks  have opposite signs for opposite sweep orientations.  The right-side diagram and plots refer to a case in which  the orientation of spin remains the same (either $+\rm{\pi}/2$ or $-\rm{\pi}/2$), Therefore, the peaks have the same sign for both sweep orientations.}.
\label{fig:Spin-texture}
\end{figure*}
\subsection{Chiral domain walls}
A spin texture for domain walls (see supplemental material in \cite{Liu2017}), which would explain our results, is sketched in Fig.~\ref{fig:Spin-texture}. One domain (oriented along $\theta=\rm{\pi}$) is located at the center and another domain with opposite polarity ($\theta=0$) at the periphery. [In the convention used here \cite{Liu2017}, $\theta$ is the angle between the $x$-axis and a pair of parallel spins of the unit cell.] In the (more or less thick) wall separating these two domains, spins rotate smoothly and concomitantly in the $x-y$ plane. The texture along $x$-axis is such that at the center of the domain wall, the adopted configuration has an orientation perpendicular to the two domains.  Fig.~\ref{fig:Spin-texture}b shows different versions of the same structure with a narrower wall. One can see that the two possible configurations are  $+\rm{\pi}/2$ and $-\rm{\pi}/2$. This would correspond to an either clockwise or anticlockwise rotation of spins depending on the specific domain configuration at the center and the periphery. Note that domain walls of this type, with in-plane rotation of two possible signs, follow directly from the hierarchy of scales discussed in \cite{Liu2017}, in which the Dzyaloshinskii-Moriya (DM) interaction is much stronger than an in-plane two-fold anisotropy. The origin of the two-fold anisotropy will be discussed in future work.

We note that a study using Magneto-Optical Kerr Effect (MOKE) microscopy \cite{Higo2018} detected oppositely aligned domains in the multidoamin regime at small magnetic fields. The domains were found to extend over tens of microns. However, the fine structure of the walls separating these domains  \cite{Liu2017} could not be resolved in this study.

Such a texture would provide a natural explanation for the transverse magnetization (TM) and the planar Hall effect observed in regime II. The in-plane tilt of spins (and the magnetic octupole \cite{Suzuki2017}) would generate a magnetic field perpendicular to and electric field parallel to the orientation of the applied magnetic field. This is the origin of the transverse magnetization and planar Hall effect. The angle-dependent study of the AHE \cite{Li2018} has established that the orientation of the electric field associated with anomalous Hall effect is set by the orientation of spins (and not the crystal axes). Therefore, the $\rm{\pi}/2$ spin configuration in the center of domain wall would naturally gives rise to an electric field perpendicular to those generated by the $\theta=0$ and $\theta=\rm{\pi}$ domains.

In this picture, the sign of the signals reflects the chirality of the domain wall. Consider a hysteresis loop with the magnetic field swept from  a $\theta=\rm{\pi}$ single-domain  to  another $\theta=0$ single domain regime and then back to the original $\theta=\rm{\pi}$ single-domain (Fig.~\ref{fig:Spin-texture}c). If during this sequence, for both sweeping orientations, the spin configuration inside the domain walls remains the same (either +$\rm{\pi}/2$ or -$\rm{\pi}/2$), then the PHE and the TM signals will be even (symmetric) in field. On the other hand, if what remains fixed is the sense of the rotation  (clockwise or anticlockwise), then the signals will be odd (or asymmetric) in field, because the spin configuration inside the domain wall will be opposite during the two sweeps.

\begin{figure*}
\centering
\includegraphics[width=16cm]{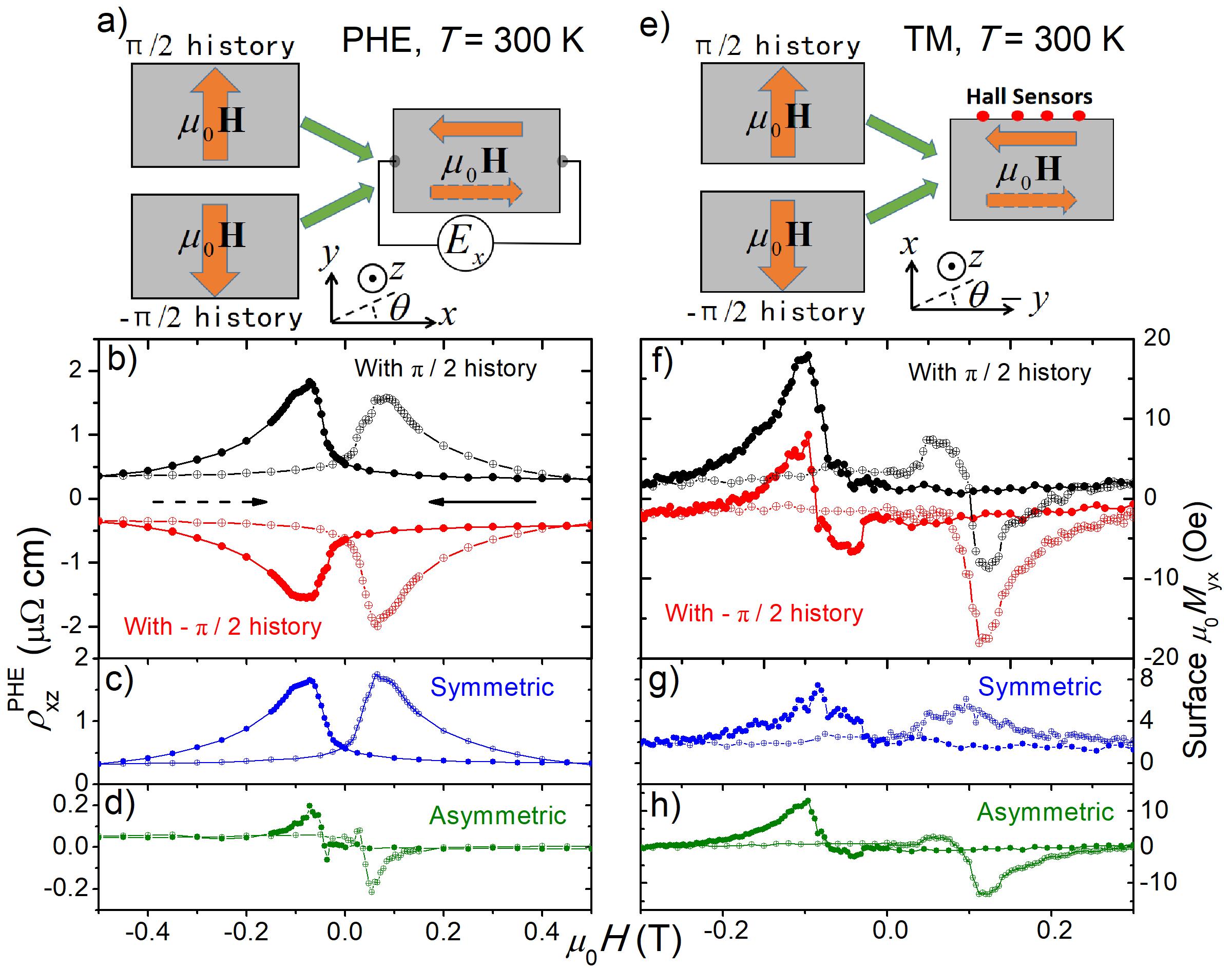}
\caption{\textbf{Dependence of planar Hall effect and transverse magnetization on prior history:} (a) The experimental protocol: Measurements were performed in identical  configurations but following different prior histories. In both cases, the applied magnetic field and the measured electric field were both along the x-axis and the magnetic field was swept from $-x$ (that is $\theta=\rm{\pi}$) to $+x$.  Before the measurement, however, in one case the  magnetic field was rotated from $-\rm{\pi}/2$ towards  $\rm{\pi}$, but in another from $+\rm{\pi}/2$ towards $\rm{\pi}$.  (b) The  planar Hall data for the two measurements. The prior history determines the sign of the observed signal. (c) The  difference between the two data sets shown in (b) (The symmetric component). (d) The sum of the two data sets shown in (b) (the asymmetric response). (e) The experimental protocol for measuring transverse magnetization, similar to (a).  (f)  Transverse magnetization data for the two measurements. (g) The difference between the two measurements in (f) (The symmetric component). (h) The sum of the two measurements in (f) (the asymmetric component). The closed and open symbols refer to opposite field sweep orientations, marked by solid and dotted arrows shown in (b).
\label{fig:Memory-effect}}
\end{figure*}

\begin{figure}
\includegraphics[width=9cm]{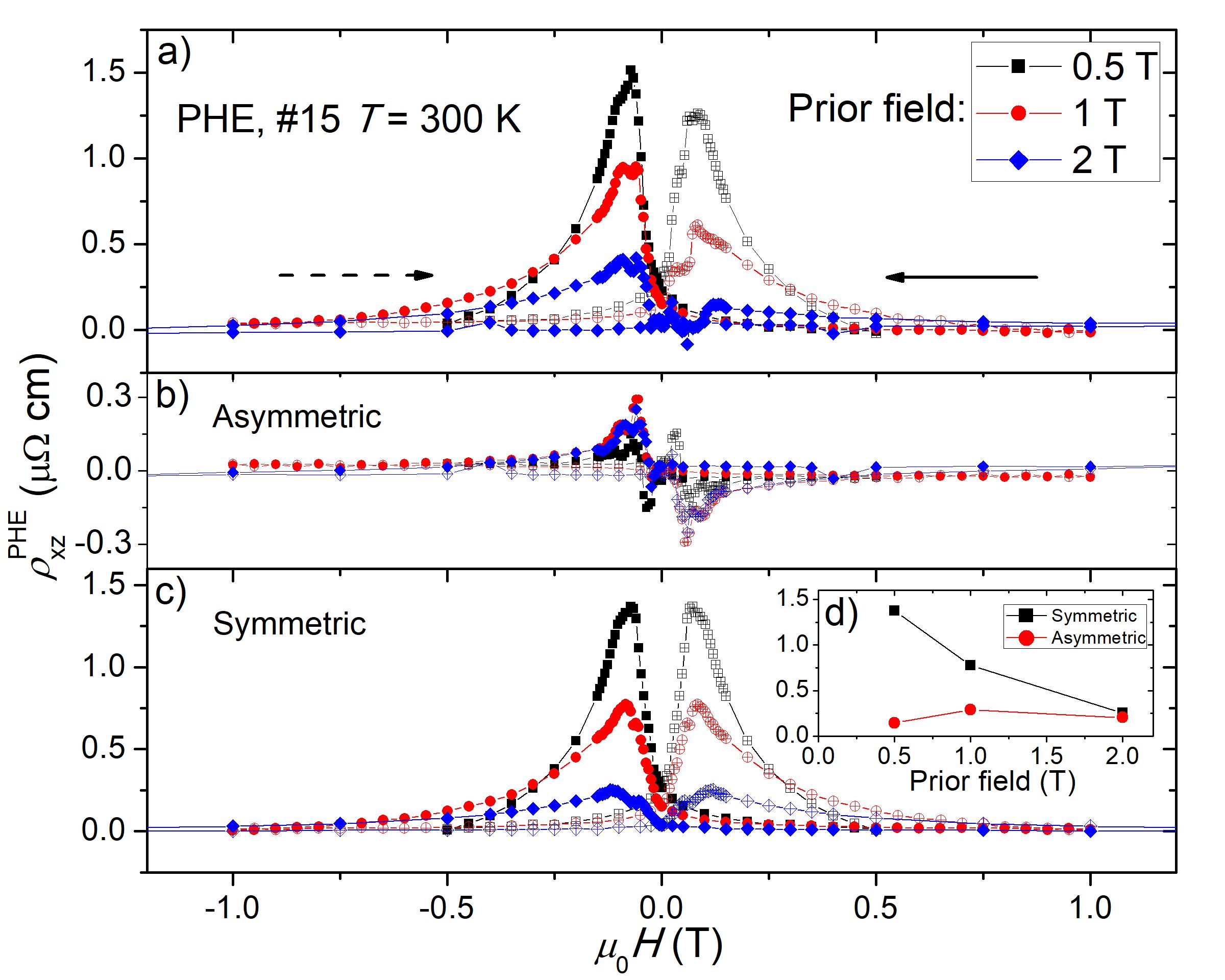}
\caption{\textbf{Evolution of the PHE signal with the amplitude of the prior magnetic field:} (a) Planar Hall effect measured after cycling and rotating the magnetic field at 0.5 T, 1 T and 2 T. (b) and (c) Asymmetrical and symmetrical components of planar Hall effect, extracted from (a). Inset in (c) shows the evolution of the magnitudes of the two components. The symmetric component steadily decreases with the increasing prior magnetic field, but the asymmetric component does not. The closed and open symbols refer to opposite field sweep orientations, marked by solid and dotted arrows shown in (a).
\label{fig:Field-dependet-PHE}}
\end{figure}

\subsection{Domain walls have a memory }
Keeping this in mind, let us turn our attention to another outcome of this study, a memory effect. The experimental protocol is defined in (fig.~\ref{fig:Memory-effect}a). We performed the measurement twice for identical configurations, but with different prior histories. The measurement consisted in sweeping the magnetic field oriented along $x$-axis from 0.5 T to -0.5 T and back. This corresponds to switching domains from $\theta=0$ to $\theta=\rm{\pi}$ configurations and back to the starting point. The measurement was preceded in the first case by a field rotation from $+\rm{\pi}/2$ to 0 and in the second case, by a rotation from $-\rm{\pi}/2$ to 0. As one can see in fig.~\ref{fig:Memory-effect}b), the results are strikingly different.  In the first case the PHE signals are positive, in the second are negative. We note that this is a phenomenon belonging to the category dubbed memory of direction \cite{Keim2018}. By subtracting the two sets of data or adding them, one can extract the symmetric (fig.~\ref{fig:Memory-effect}c) and the asymmetric (fig.~\ref{fig:Memory-effect}d) components of the PHE signal. The symmetric part is seven times larger than the asymmetrical part.

Note the small gap seen between the two sets of $\rho^{\rm{PHE}}_{xz}$ data obtained with two different prior histories in Fig. \ref{fig:Memory-effect}b. It arises because we have assumed an identical offset for both sets of data. This offset is caused by an unavoidable misalignment between lateral contacts. The difference between the two sets of data obtained with different prior histories implies that history affects the offset too.

The transverse magnetization displays also a similar memory (See the protocol defined in Fig.~\ref{fig:Memory-effect}e and the data shown in Fig.~\ref{fig:Memory-effect}f-h). One can see, however, that in this case, the main component is asymmetric, which is three times larger than the symmetric one.

\section{Discussion}
Recalling that PHE is a bulk effect, we conclude that the orientation of the spins inside the walls is mainly set by the past history. On the other hand, in the case of transverse magnetization at the surface, the spin orientation mainly depends on the sign of the magnetic field and the rotation orientation is less affected by the prior history. This raises an obvious question: Where does the system stock the information regarding the previous orientation of the magnetic field?

A plausible answer to this question is provided by the scenario sketched above. When  the magnetic field is oriented along $\theta=+\rm{\pi}/2$, at the end of a ($-y\leftrightarrow+y$) hysteretic loop, the sample is practically single-domain with $\rm{\pi}/2$ spin configuration. In principle rotating from $\rm{\pi}/2$ to $0$ before the measurement would change the spin configuration of the whole sample from $\rm{\pi}/2$ to $0$. However, if residual domains remain stuck in the $\rm{\pi}/2$ configuration, they will set the spin configuration of the domain walls along $\rm{\pi}/2$.  If this is the case, then one would expect to see a dependence of the memory effect on the strength of the prior magnetic field. The larger the magnetic field at which the ($\rm{\pi}/2$ to $0$) rotation takes place before  the measurement, the smaller the fraction of the domains which had stayed in place and the smaller their role in setting the chirality.

As seen in Fig.~\ref{fig:Field-dependet-PHE}, this is indeed the case.  We measured PHE  after cycling and rotating the magnetic field at B= 0.5 T, 1 T, 2 T. One can see that the magnitude of the PHE and in its symmetric component steadily decreases. This implies that the symmetric component of the PHE set by the chirality of the wall is promoted by the presence of minority domains, whose population decrease with increasing magnetic field. The asymmetric component, on other hand, does not show significant evolution with magnetic field.

If domain walls with opposite chiralities were evenly distributed in the sample£¬ no PHE or transverse magnetization signal would have been observed. This is not the case. The dominance of a  symmetric and history-dependent component in the PHE signal implies that deep inside the sample, minority domains set the chirality of the domain wall. The dominance of the asymmetric and history-independent component in surface transverse magnetization indicates that wall spin orientation at surface is principally set by the orientation of the magnetic field with only a minor role proposed by the minority domains.
We note that the domain wall spin texture proposed here can also generate a topological Hall response as reported previously \cite{Li2018}, provided that we assume an additional off-plane tilt of spins residing inside the domain walls. Indeed, if the unit vector of magnetization has a finite z dependence ($\frac{\partial \overrightarrow{n}}{\partial z}\neq 0 $), then combined with the finite $\frac{\partial \overrightarrow{n}}{\partial r }$, it generates an axially oriented emergent magnetic field (B$_{\theta}\neq 0$ in cylindrical coordinates) \cite{Everschor2014} and the skyrmionic number will be finite, producing real-space Berry curvature. Such an assumption would not alter the conclusions drawn above. Yet, it is not necessary for explaining the observations reported in the present study.

Heating the sample above $T_{\rm{N}}=$ 420 K would presumably erase all history dependence. It would be interesting to compare field-cooled and zero-field-cooled behaviors across the transition temperature in future experiments combining  a furnace and a magnet.

In summary we put under scrutiny a narrow field window in which there are multiple magnetic domains in Mn$_3$Sn and found that in this regime, one can observe a planar Hall and planer Nernst effect  as well as transverse magnetization. These observations can be explained by a specific spin texture for domain walls where spins rotate in the pseudo-Kagom\'e plane. The choice of clockwise or anti-clockwise rotation can be controlled by the prior magnetic history of the sample, providing a new platform for memory formation.

\section{Methods}
\textbf{Sample preparation and Transport measurements:} Mn$_3$Sn single crystals with a typical size in the range of centimeter were grown by the vertical Bridgman technique [24]. They were cut to desired dimensions by a wire saw. All transport experiments were performed in a commercial measurement system (Quantum Design PPMS), using the Horizontal Rotator Option. Hall resistivity was measured by a standard four-probe method using a current source and a DC-nanovoltmeter. Two Chromel- Constantan (type E) thermocouples were employed to measure the temperature difference in the case of Nernst
measurements.

\textbf{magnetization:} Bulk magnetization was measured using a vibrating sample magnetometer (VSM) method. For surface magnetization measurements, we employed an array of Hall sensors based on high-mobility AlGaAs/GaAs heterostructure; The density of the two-dimensional electron gas (2DEG) was n = 2.5 $\times$ 10$^11$cm$^{-2}$ (300 K) and it was located 160 nm below the surface. The device was fabricated using electron beam lithography and 250 V argon ions to define the mesa. Supplementary Figure 1 shows an array of ten sensors each $5\times5 \mu \rm{m}^2$ square with a 100 $\mu$m interval between two neighboring sensors \cite{Collignon2017}. Attaching the device to the surface of the sample, the local magnetic field was determined by measuring the Hall resistivity of the sensor using an AC current source and a lock-in amplifier.

\textbf{Data availability}

The data that support the findings of this study are available from the corresponding author upon reasonable request.

\noindent
* \verb|zengwei.zhu@hust.edu.cn|\\
* \verb|kamran.behnia@espci.fr|\\

\textbf{Acknowledgements-} This work was supported by the National Science Foundation of China (Grant No. 51861135104 and 11574097), the National Key Research and Development Program of China (Grant No.2016YFA0401704), the Fundamental Research Funds for the Central Universities (Grant No. 2019kfyXMBZ071) and by Agence Nationale de la Recherche (ANR-18-CE92-0020-01). ZZ was supported by the 1000 Youth Talents Plan. KB was supported by China High-end foreign expert program and Fonds-ESPCI-Paris. LB was supported by the US National Science Foundation Materials Theory program, grant number DMR1818533.  XL acknowledges a PhD scholarship by the China Scholarship Council(CSC).

\textbf{Author Contributions:} Z.Z. and K.B. supervised the research project. X. L prepared the samples. X. L. carried out experiments with helps from C. C., L. X., H. Z. and B. F.. A. C, U. G, and D. M.made the 2DEG Hall sensors used in the experiments. L.B. provided theoretical background. X. L, Z.Z. and K.B. wrote the manuscript and all authors commented on the manuscript.

\textbf{Competing interests:} The authors declare no competing financial interests.
\clearpage
\renewcommand{\thesection}{Supplementary Note \arabic{section}}
\renewcommand{\tablename}{}
\renewcommand{\thetable}{Supplementary table \arabic{table}}
\renewcommand{\thetable}{Supplementary table \arabic{table}}
\renewcommand{\thefigure}{Supplementary Figure \arabic{figure}}
\renewcommand{\figurename}{}

\renewcommand{\theequation}{Supplementary Eq. \arabic{equation}}

\setcounter{section}{0}
\setcounter{figure}{0}
\setcounter{table}{0}
\setcounter{equation}{0}
{\large\bf Supplemental Material for Chiral domain walls of Mn$_3$Sn and their memory}
{\large\bf by X. Li et al.,}
\setcounter{figure}{0}
\section{Experimental methods}
\begin{figure}[htbp]
\includegraphics[width=8.5cm]{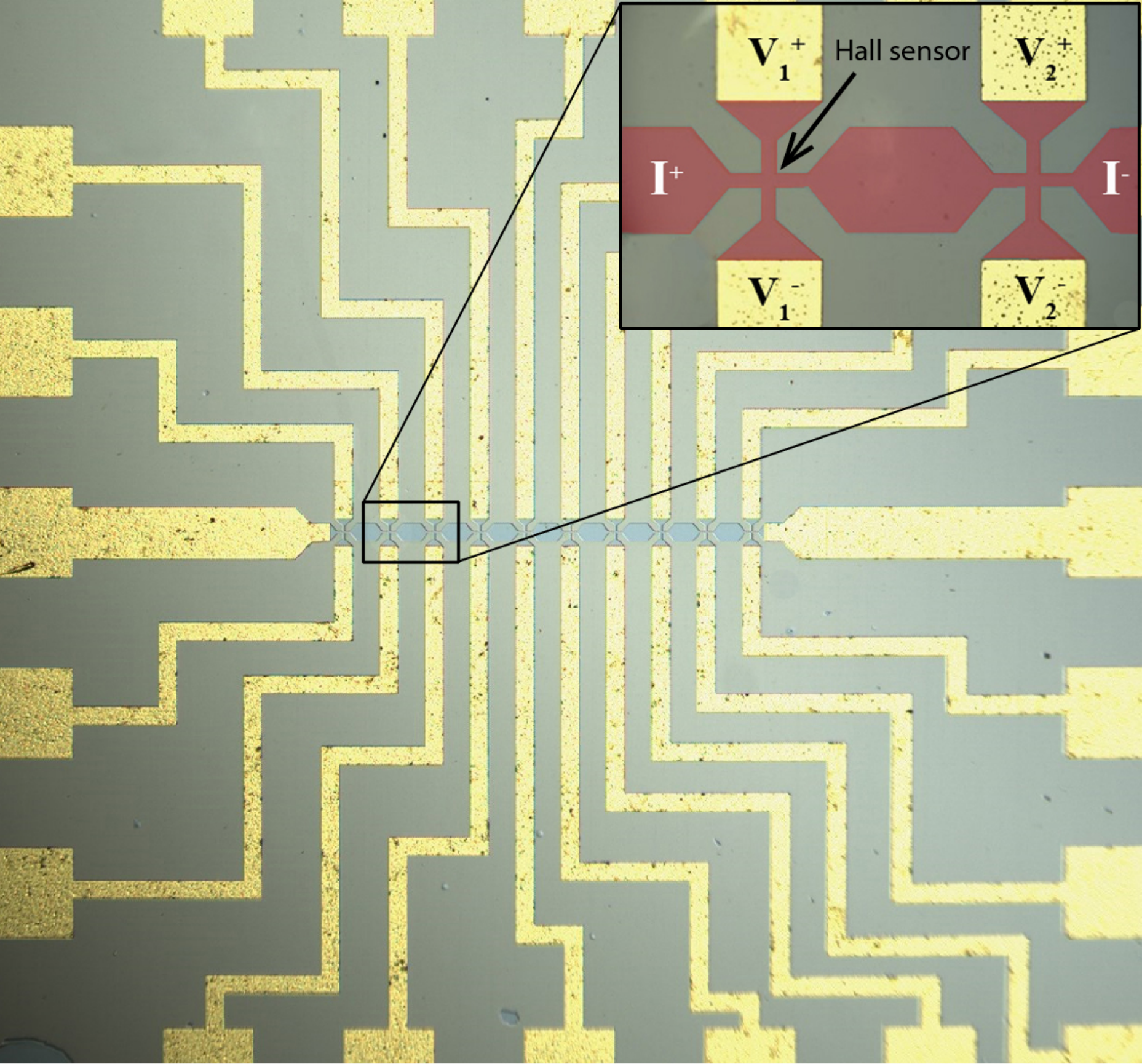}
\caption{Micron-size 2DEG Hall sensors: every device is consist of 10 Hall sensors, detecting the local field perpendicular to the plane.}
\label{S1}
\end{figure}

For surface magnetization  measurements, we employed an array of Hall sensors based on high-mobility AlGaAs/GaAs heterostructure; The density of the two-dimensional electron gas (2DEG) was $n=2.5\times10^{11}$cm$^{-2}$ (300K) and it was located 160 nm below the surface. The device was fabricated using electron beam lithography and 250 V argon ions to define the mesa. ~\ref{S1}  shows an array of ten sensors each $5\times 5 \mu$m$^{2}$ square with a 100 $\mu$m interval between two  neighboring sensors \cite{Collignon2017}. Attaching the device to the surface of the sample, the local magnetic field was determined by measuring the Hall resistivity of the sensor using an AC current source and a lock-in amplifier.

\section{Sample details}
For different experiments, six samples were used in this work with different aspect ratios ($l_y/l_x$) and cross-section shapes. The list of samples is given in \ref{tableS1}. Samples \#5 and \#13-1 have square cross-section samples and aspect ratio ($l_y/l_x$) is close to unity. Samples \#13-2, \#13-3 and \#5 have rectangular cross sections with aspect ratio ($l_y/l_x$) deviating from unity; The sample dubbed \#triangle,  has a equilateral triangle as cross section, with three sides along y axis.

\begin{table}[htb]
\setlength{\tabcolsep}{0.3cm}{
\begin{tabular}{|c|c|c|c|c|}
\hline
 &$l_{x}$(mm)&$l_{y}$(mm)&$l_{z}$(mm)&$l_y/l_x$ \\
\hline
\#5&0.5&0.6&2&1.2 \\
\hline
\#13-1&0.5&0.54&0.66&1.08 \\
\hline
\#13-2&0.28&0.54&0.66&1.93 \\
\hline
\#13-3&0.16&0.54&0.66&3.38 \\
\hline
\#15&1&0.2&1.8&0.2 \\
\hline
\#triangle&0.63&0.73&1.8&1.16 \\
\hline
\end{tabular}}
\caption{Size and the aspect ratio $l_y/l_x$ of six different samples used in this work.}
\label{tableS1}
\end{table}

\section{Planar Hall effect (PHE) in different samples}
\begin{figure}
\includegraphics[width=9cm]{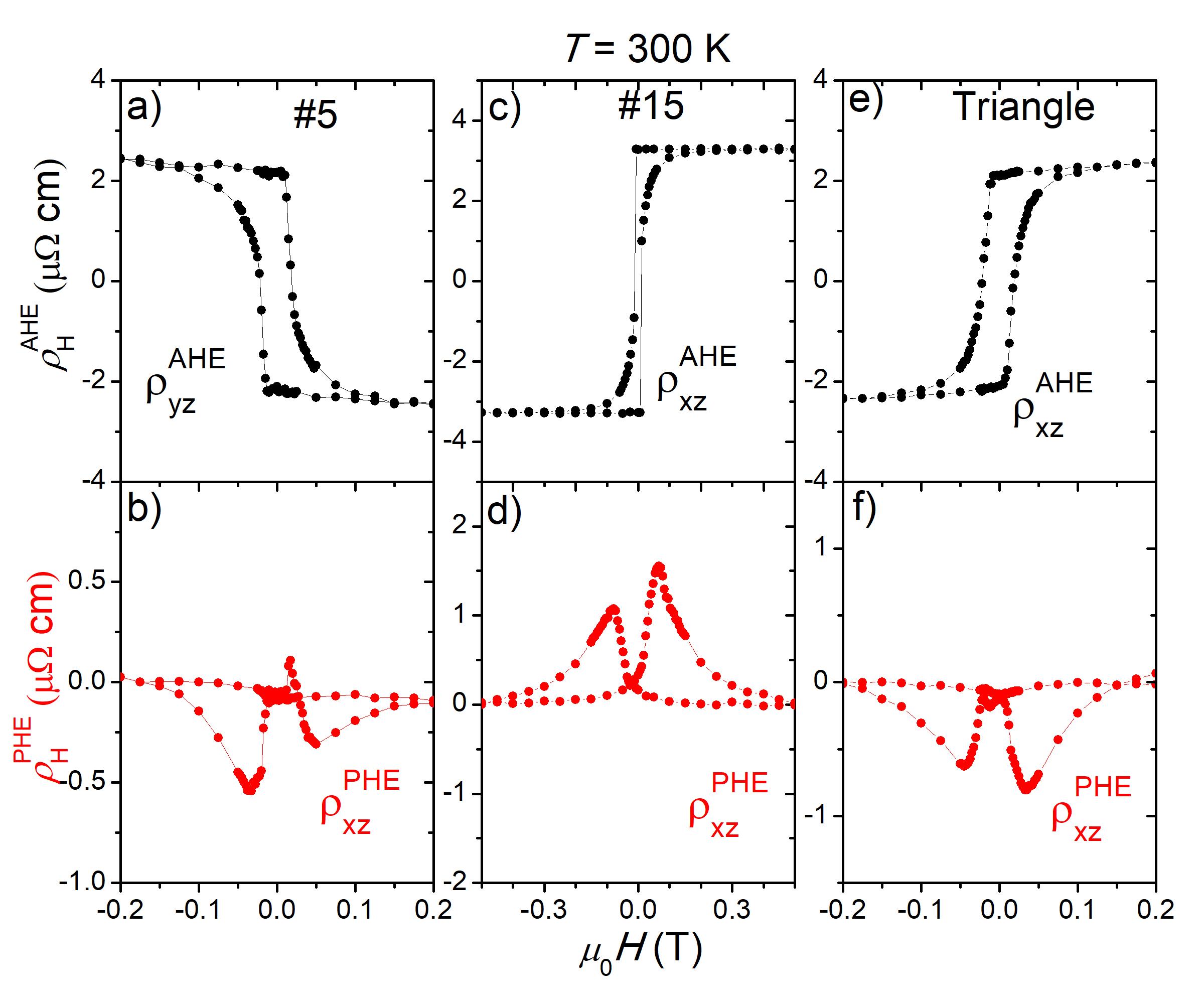}
\caption{Anomalous and planar Hall resistivity in the square sample \#5, rectangular sample \#15 and the triangle sample.}
\label{S2}
\end{figure}

The existence of the planar Hall effect was reproduced in more than three samples and with different set-ups and  cross-sections (~\ref{S2}). ~\ref{S2}a and ~\ref{S2}b represent anomalous and planar Hall resistivity in sample \#5 with the magnetic field along x axis, and the electric field measured simultaneously along both $x$- and $y$-axes.  ~\ref{S2}c and ~\ref{S2}d show anomalous and planar Hall resistivity in sample \#15, the one in which the  Nernst data was shown in the main text. The electric field $E_x$  was measured for  two different orientations of the magnetic field. ~\ref{S2}e and ~\ref{S2}f show the Hall data in a sample with triangle cross-section. The planar Hall effect is present in all these three samples and  the ratio ($\rho^{\rm{PHE}}$/$\rho^{\rm{AHE}}$) is always between 0.3 to 0.4. It's worth noting that the width of the regime II is different and depended on the aspect ratio $l_y/l_x$ \cite{Li2018}.

\section{Extraction of topological Hall effect (THE) and topological Nernst effect (TNE)}
\begin{figure}
\includegraphics[width=9cm]{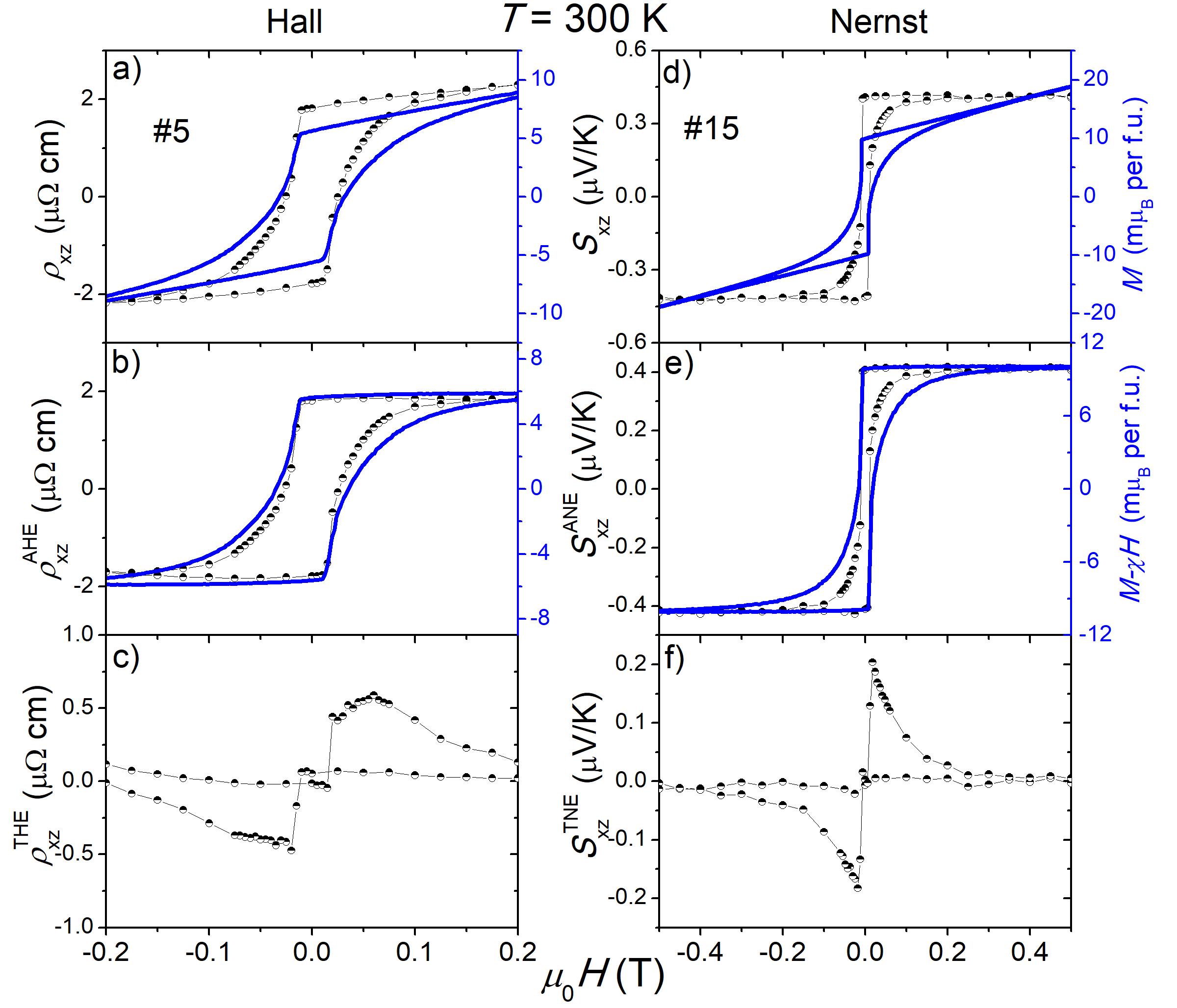}
\caption{(a) and (d) THe Hall and Nernst signals compared with the magnetization. (b) and (e) Comparison of the anomalous Hall and Nernst response with the magnetization subtracted the high field slope. (c) and (f) The topological Hall and Nernst response subtracted from (b) and (e).}
\label{S3}
\end{figure}

 ~\ref{S3}a and ~\ref{S3}d compares the hysteretic loops of the Hall and the Nernst response with the magnetization. ~\ref{S3}b and ~\ref{S3}e show the  same comparison between normalized signals (after subtracting the high-field slope).
 One can see that the two responses do not scale with each other in regime II. ~\ref{S3}c and ~\ref{S3}f show  $\rho^{\rm{THE}}_{xz}(\mu_{0}H)=\rho^{\rm{A}}_{xz}(\mu_{0}H)-C(M(\mu_{0}H)-\mu_{0}H\chi)$ and  $S^{\rm{TNE}}_{xz}(\mu_{0}H)= S^{\rm{A}}_{xz}(\mu_{0}H)-C^*(M(\mu_{0}H)-\mu_{0}H\chi)$ , where $\chi$ is the high-field susceptibility (the slope of the magnetization outside the hysteresis loop), $C=\rho^{\rm{A}}_{xz}(\mu_{0}H=0T)/M(\mu_{0}H=0T)$ and $C^*=S^{\rm{A}}_{xz}(\mu_{0}H=0T)/M(\mu_{0}H=0T)$ are constant two fitting constants.\\
\section{Temperature dependence of PHE and PNE}
\begin{figure}
\includegraphics[width=9cm]{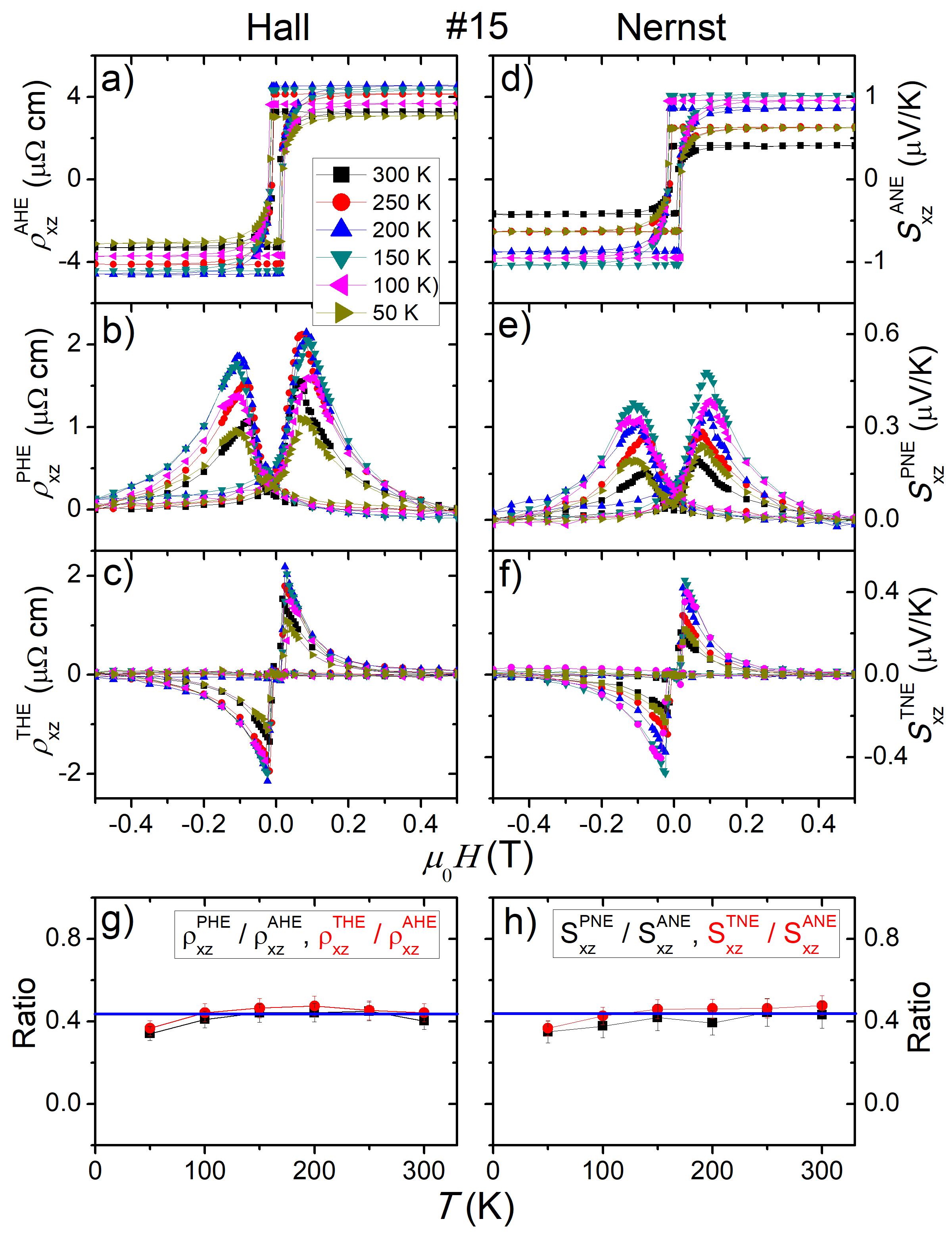}
\caption{ Three types of Hall response (a-c) and Nernst response (d-f) with the temperature from 300K to 50K. (g) and (h) Temperature dependent ratio of $\rho^{\rm{PHE}}_{xz}$/$\rho^{\rm{AHE}}_{xz}$(black square left) and $\rho^{\rm{PHE}}_{xz}$/$\rho^{\rm{AHE}}_{xz}$ (red circle left), $S^{\rm{PNE}}_{xz}$/$S^{\rm{ANE}}_{xz}$(black square right) and $S^{\rm{PNE}}_{xz}$/$S^{\rm{ANE}}_{xz}$ (red circle right). All the ratios keep a constant near 0.4.}\label{S4}
\end{figure}

 ~\ref{S4}, shows the evolution of PHE and PNE as the temperature changes from 300 K to 50 K. In the whole temperature range studied,  the ratios  of both PHE(THE) and PNE(TNE) to AHE and ANE remain constant.

\section{Spin texture with rectangular boundaries}
\begin{figure*}
\centering
\includegraphics[width=18cm]{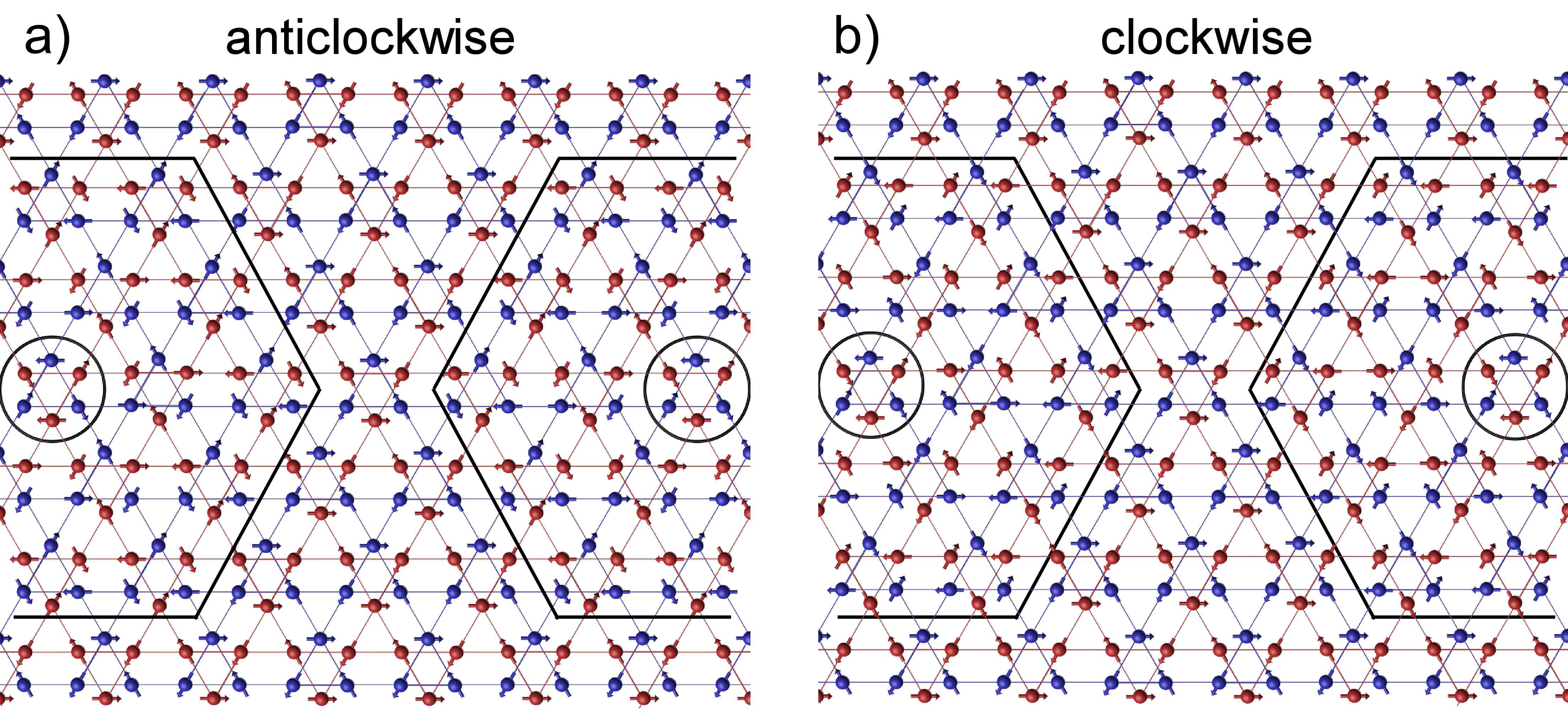}
\caption{Domain wall with clockwise (a) or anticlockwise (b) rotation with rectangular boundaries.}
\label{S5}
\end{figure*}

~\ref{S5}, shows two domains separated by a domain wall in a rectangular configuration\cite{Liu2017}.

\begin{thebibliography}{99}
\bibitem{Getzlaff} Getzlaff M. Fundamentals of Magnetism. Springer Science \& Business Media (2007).
\bibitem{Tetienne} Tetienne, J. P. \textit{et al.} The nature of domain walls in ultrathin ferromagnets revealed by scanning nanomagnetometry. Nature Commun. \textbf{6}, 6733 (2015).
\bibitem{Cheng} Cheng, R. \textit{et al.} Magnetic domain wall Skyrmions. Phys. Rev. B \textbf{99}, 184412 (2019).
\bibitem{Stamps} Stamps, R. L. \textit{et al}. J. Phys. D: Appl. Phys. \textbf{47} 333001 (2014).
\bibitem{Nakatsuji}  Nakatsuji, S., Kiyohara, N. \& Higo, T. Large anomalous Hall effect in a non-collinear antiferromagnet at room temperature. Nature \textbf{527}, 212 (2015).
\bibitem{Nayak}  Nayak, A. K. \textit{et al.} Large anomalous Hall effect driven by a nonvanishing Berry curvature in the noncolinear antiferromagnet Mn$_{3}$Ge. Sci. Adv. \textbf{2:} e1501870 (2016).
\bibitem{Kiyohara} Kiyohara, N., Tomita, T. \& Nakatsuji, S. Giant Anomalous Hall Effect in the Chiral Antiferromagnet Mn$_{3}$Ge. Phys. Rev. Appl. \textbf{5}, 064009 (2016).
\bibitem{Zimmer} Zimmer, G. J. \&  Kr\'{e}n, E. Investigation of the Magnetic Phase Transformation in Mn$_{3}$Sn. AIP Conf. Proceed.  \textbf{5},  513 (1972).
\bibitem{Tomiyoshi1982} Tomiyoshi, S. Polarized neutron diffraction study of the Spin Structure of Mn$_3$Sn. J. Phys. Soc. Jpn. \textbf{51}, 803 (1982).
\bibitem{Tomiyoshi1982b} Tomiyoshi, S. \& Yamaguchi, Y. Magnetic Structure and Weak Ferromagnetism of Mn$_{3}$Sn Studied by Polarized Neutron Diffraction. J. Phys. Soc. Jpn. \textbf{51}, 2478 (1982).
\bibitem{Yang2017} Yang, H. \textit{et al.} Topological Weyl semimetals in the chiral antiferromagnetic materials Mn$_{3}$Ge and Mn$_{3}$Sn. New J. Phys. \textbf{19}  015008(2017).
\bibitem{Chen2014} Chen, H., Niu, Q. \& MacDonald, A. H. Anomalous Hall Effect Arising from Noncollinear Antiferromagnetism. Phys. Rev. Lett. \textbf{112}, 017205 (2014).
\bibitem{Kubler2014} K\"{u}bler, J. \& Felser, C. Non-collinear antiferromagnets and the anomalous Hall effect. Europhys. Lett. \textbf{108}, 67001 (2014).
\bibitem{Ikhlas2017} Ikhlas, M. \textit{et al.} Large anomalous Nernst effect at room temperature in a chiral antiferromagnet. Nat. Phys. \textbf{13}, 1085 (2017).
\bibitem{Li2017} Li, X. \textit{et al.} Anomalous Nernst and Righi-Leduc effects in Mn$_{3}$Sn: Berry curvature and entropy flow. Phys. Rev. Lett. \textbf{119}, 056601 (2017).
\bibitem{Higo2018} Higo, T. \textit{et al.} Large magneto-optical Kerr effect and imaging of magnetic octupole domains in an antiferromagnetic metal. Nature Photonics \textbf{12}, 73 (2018).
\bibitem{Xu2018}  Xu, L. \textit{et al.} Finite-temperature violation of the anomalous transverse Wiedemann-Franz law in absence of inelastic scattering. arXiv:1812.04339 (2018).  Preprint at https://arxiv.org/abs/1812.04339 (2018)
\bibitem{Balk2019} Balk, A. L. \textit{et al.} Comparing the anomalous Hall effect and the magneto-optical Kerr effect through antiferromagnetic phase transitions in Mn$_3$Sn.
Appl. Phys. Lett. \textbf{114}, 032401 (2019).
\bibitem{Wuttke2019} Wuttke, C. \textit{et al.}  Berry curvature unravelled by the Nernst effect. arXiv:1902.01647 (2019). Preprint at https://arxiv.org/abs/1902.01647 (2019)
\bibitem{Sugii2019} Sugii, K. \textit{et al.} Anomalous thermal Hall effect in the topological antiferromagnetic state. arXiv:1902.06601 (2019). Preprint at https://arxiv.org/abs/1902.06601 (2019)

\bibitem{Zhang2017}Zhang, Y. \textit{et al.} Strong anisotropic anomalous Hall effect and spin Hall effect in the chiral antiferromagnetic compounds Mn$_3$X (X=Ge, Sn, Ga, Ir, Rh, and Pt). Phys. Rev. B \textbf{95}, 075128 (2017).

\bibitem{Smejkal2018} Smejkal, L., Mokrousov, Y., Yan, B. \& MacDonald, A. H. Topological antiferromagnetic spintronics. Nat. Phys. \textbf{14}, 242 (2018).


\bibitem{Liu2017} Liu, J. \& Balents, L. Anomalous Hall Effect and Topological Defects in Antiferromagnetic Weyl Semimetals: Mn$_{3}$Sn/Ge. Phys. Rev. Lett. \textbf{119}, 087202 (2017).
\bibitem{Li2018} Li X. \textit{et al.} Momentum-space and real-space Berry curvatures in Mn$_{3}$Sn. SciPost Phys. \textbf{5}, 063 (2018).
\bibitem{Behnia2000} Behnia K.,  Capan C., Mailly D. \&  Etienne B. Internal avalanches in a pile of superconducting vortices. Phys. Rev. B \textbf{61}, R3815 (2000).
\bibitem{Collignon2017} Collignon, C. \textit{et al.} Superfluid density and carrier concentration across a superconducting dome: The case of strontium titanate. Phys. Rev. B \textbf{96}, 224506 (2017).

\bibitem {Keim2018} Keim, N. C., Paulsen, J. , Zeravcic, Z., Sastry, S. \& Nagel S. R. Memory formation in matter. arXiv:1810.08587
Preprint at https://arxiv.org/abs/1810.08587 (2018)

\bibitem{Suzuki2017} Suzuki, M.T., Koretsune, T., Ochi, M. \& Arita, R. Cluster multipole theory for anomalous Hall effect in antiferromagnets. Phys. Rev. B \textbf{95}, 094406 (2017).

\bibitem{Everschor2014} Everschor-Sitte, K. \& Sitte, M. Real-space Berry phases: Skyrmion soccer. J. Appl. Phys. \textbf{115}, 172602 (2014).
\end{thebibliography}
\end{document}